\documentclass{eptcs}
\usepackage[utf8]{inputenc}
\usepackage{microtype}
\usepackage{listings}
\usepackage{amsmath}
\usepackage[figuresleft]{rotating}
\usepackage{booktabs}

\hypersetup{colorlinks,urlcolor=blue,citecolor=blue,linkcolor=blue}

\newcommand\ellipse[1][]{\text{\fcolorbox{black}{black!10}{\vphantom{x}{\itshape#1}\dots}}}

\usepackage[backend=bibtex,firstinits,sorting=nyt]{biblatex}
\bibliography{wflp2016}

\usepackage[scaled=0.85]{helvet}

\usepackage{stmaryrd}
\usepackage{xcolor}
\colorlet{new}{black}
\newcommand\NEW[1]{{\color{new}#1}}

\newdimen\Commentwidth
\newcommand*{\Comment}[1]{\hfill\makebox[\Commentwidth][l]{\sffamily\slshape//\enspace#1}}%
\title{A Practical Study of Control in Objected-Oriented--Functional--Logic Programming with Paisley}
\author{Baltasar Trancón y Widemann \institute{Ilmenau University of Technology, DE} \email{baltasar.trancon@tu-ilmenau.de} \and Markus Lepper \institute{\texttt{<semantics\kern1pt/>} GmbH, DE}}

\def\titlerunning{Control for Paisley}
\def\authorrunning{Trancón y Widemann \& Lepper}

\begin{document}
\maketitle

\begin{abstract}
  Paisley is an extensible lightweight embedded domain-specific
  language for nondeterministic pattern matching in Java.  Using
  simple APIs and programming idioms, it brings the power of
  functional--logic processing of arbitrary data objects to the Java
  platform, without constraining the underlying object-oriented
  semantics.  Here we present an extension to the Paisley framework
  that adds pattern-based control flow.  It exploits recent additions
  to the Java language, namely functional interfaces and lambda
  expressions, for an explicit and transparent continuation-passing
  style approach to control.  We evaluate the practical impact of the
  novel features on a real-world case study that reengineers a
  third-party open-source project to use Paisley in place of
  conventional object-oriented data query idioms.  We find the
  approach viable for incremental refactoring of legacy code,
  with significant qualitative improvements regarding separation of
  concerns, clarity and intentionality, thus making for easier code
  understanding, testing and debugging.
\end{abstract}


\section{Introduction}

The object-oriented paradigm, in its pure form, suffers from strongly
asymmetric expressivity with respect to structured data.  On the one
hand, \emph{constructors} and \emph{factory methods} allow for a
term-like notation for the construction of data.  On the other hand,
the corresponding dedicated tools for the query and deconstruction of
data, namely dynamic \emph{type tests}, \emph{casts} and \emph{getter
  methods}, are not only very different in appearance, but also
markedly less expressive and compositional.  Programming style
patterns and idioms such as \emph{iterators} and \emph{visitors} have
been developed to amend the issue, but suffer from restricted
applicability and heavy impact on code structure, resulting in
imprecise and clumsy practical usage.

By contrast, declarative languages typically support a notation that
is directly inverse to, and hence as clear, lightweight, precise and
expressive as, data construction: \emph{pattern matching}.  The
denotational semantics of patterns rely on the invertible algebraic
structure of declarative data models.  Hence they do not carry over to
arbitrary object-oriented data models, whose basic principles such as
transcendental identity, mutable state and data abstraction are
fundamentally at odds.

Multi-paradigm languages such as Scala\,\cite{scala} have demontrated
that object-oriented programming can benefit greatly from an approach
to pattern matching that is more operational and hence adequate to
imperative program semantics.  By contrast, approaches such as
JMatch\,\cite{jmatch}, which impose declarative pseudo-semantics on
objects and imperative code, have never successfully reached the
mainstream.

We have designed Paisley\,\cite{atps2012} as a lightweight embedded
domain-specific language, that is a library of classes and idioms, for
pattern matching in Java.  Paisley integrates closely with
the imperative object-oriented paradigm, even closer than Scala's
built-in patterns.  It supports user extensions,
combinatorial abstractions, nondeterminism by backtracking,
encapsulated search, and metaprogramming, without requiring changes to
either the Java host language or the code of the data object models
to be queried.

In past papers, we have demonstrated how to use Paisley to sanitize
legacy query interfaces~\cite{icmt2012}, how to combine nondeterminism
and search-plan metaprogramming to solve logical
puzzles~\cite{wflp2013}, and how to express complex relational queries
in pattern (Kleene) algebra~\cite{wflp2014}, respectively.  In
summary, Paisley turns Java into an effective functional--logic
programming language, without constraining the use of features of the
object-oriented host environment.  However, the demonstrations have so
far focused on encapsulated search, and omitted a different aspect
that is ubiquitous in declarative pattern matching: the joint
specification of \emph{data and control flow} in case distinctions by
pattern matching \emph{clauses}.

The present paper fills this gap.  The Java language has adopted, in
its recent version~8, mechanisms for local functional programming that
are well-integrated with the object-oriented background.  We
demonstrate how to use these to encode the right hand sides of pattern
matching clauses as continuations, such that the ordinary Java means
for clause selection, namely conditional statements and
short-circuiting Boolean operators, can be used freely and
effectively.  Unlike for the preceding papers
\cite{wflp2013,wflp2014}, we illustrate the real-world use of this
novel expressivity not by constructing a new program for the purpose,
but instead by reengineering parts of an existing third-party
open-source project.  We discuss and evaluate the impact of our
modifications both qualitatively and quantitatively.

\section{Paisley in a Nutshell}

\begin{figure}
  \centering
  \begin{minipage}[t]{0.45\textwidth}
  \begin{lstlisting}[frame=tblr,columns=fullflexible]
public class Pair {

  private Object car, cdr;
  public Pair (Object car,
               Object cdr) {
    this.car = car;
    this.cdr = cdr;
  }
  public Object getCar() {
    return car;
  }
  public Object getCdr() {
    return cdr;
  }
  public static final Object empty;
}    
  \end{lstlisting}        
  \end{minipage}%
  \begin{minipage}[t]{0.6\textwidth}
  \begin{lstlisting}[frame=tblr,columns=fullflexible]
Pattern<Object>     isPair = isInstanceOf(Pair.class);
Motif<Pair, Object> asPair = forInstancesOf(Pair.class); 

Pattern<Object> pair (Pattern<Object> pcar,
                      Pattern<Object> pcdr) {
  return asPair.apply(car.apply(pcar)
                      .and(cdr.apply(pcdr)));
}
Motif<Object, Pair> car    = transform(Pair::getCar);


Motif<Object, Pair> cdr    = transform(Pair::getCdr);


Pattern<Object> isEmpty = eq(Pair.empty);
$\hbox{}$
  \end{lstlisting}
  \end{minipage}
  \caption{LISP lists, Java data model (left) and Paisley pattern library (right)}
  \label{fig:pair}
\end{figure}

\begin{figure}
  \begin{lstlisting}[frame=tblr,columns=fullflexible]
  list1 = new Pair(x, new Pair(y, new Pair(z, empty)));
  \end{lstlisting}
  \begin{lstlisting}[frame=tblr,columns=fullflexible]
  boolean success = false;                                                   ^\Comment{[S]}^
  if (list1 instanceof Pair) {                                               ^\Comment{[T]}^
    Pair pair1 = (Pair)list1;                                                ^\Comment{[C]}^
    Object x = pair1.getCar();                                               ^\Comment{[B,P]}^
    Object list2 = pair1.getCdr();                                           ^\Comment{[b,P]}^
    if (list2 instanceof Pair) {                                             ^\Comment{[T]}^
      Pair pair2 = (Pair)list2;                                              ^\Comment{[C]}^
      Object y = pair2.getCar();                                             ^\Comment{[B,P]}^
      Object list3 = pair2.getCdr();                                         ^\Comment{[b,P]}^
      if (list3 instanceof Pair) {                                           ^\Comment{[T]}^
        Pair pair3 = (Pair)list3;                                            ^\Comment{[C]}^
        Object z = pair3.getCar();                                           ^\Comment{[B,P]}^
        Object list4 = pair3.getCdr();                                       ^\Comment{[b,P]}^
        if (list4 == empty) {                                                ^\Comment{[T]}^
          succeed(x, y, z);                                                  ^\Comment{[R]}^
          success = true;                                                    ^\Comment{[S]}^
        }
      }
    }
  }
  if (!success)                                                              ^\Comment{[S]}^
    fail();                                                                  ^\Comment{[R]}^
  \end{lstlisting}
  \begin{lstlisting}[frame=tblr,columns=fullflexible]
  Variable<Object>  x = new Variable<>(),
                    y = new Variable<>(),
                    z = new Variable<>();                                    ^\Comment{[B]}^
  Pattern<Object> triple = pair(x, pair(y, pair(z, isEmpty)));               ^\Comment{[T,C,P,b]}^
  if (triple.match(list1))                                                   ^\Comment{[S]}^
    succeed(x.getValue(), y.getValue(), z.getValue());                       ^\Comment{[R,B]}^
  else                                                                       ^\Comment{[S]}^
    fail();                                                                  ^\Comment{[R]}^
  \end{lstlisting}
  \caption{Construction of a list (top) and its inverse operation with plain Java (center) and Paisley (bottom).  For comments see text.}
  \label{fig:xyz}
\end{figure}

As a running example, consider Figure~\ref{fig:pair} (left): a simple
data model of LISP-style universal lists, implemented in Java in
object-oriented textbook style.  It is a simple matter to denote the
construction of a list, for instance of three elements \lstinline|x|,
\lstinline|y| and \lstinline|z|.  It is however a very different
matter to use the public API to denote the inverse operation: checking
whether a given list contains exactly three elements, and if so,
extract them as \lstinline|x|, \lstinline|y| and \lstinline|z|, or
otherwise do something else.  See Figure~\ref{fig:xyz} (top and
center).  The style mandated by naïve direct use of the data model
API, and prevalent in practice, that is both in teaching and industry,
is deficient in several ways:
\begin{itemize}
\item It is low-level and of little documentation value regarding the
  intentions of the programmer.
\item It lacks compositionality; although the code structure is
  repetitive, its fragments cannot be easily reused, nor inspected,
  understood, analyzed, tested and debugged independently.
\item Its sheer verbosity leaves ample room for control and data flow
  errors.
\item It entangles concerns that should, and can, be separated.
\end{itemize}
In order to illustrate the last point, we have annotated each line of
code with the respective operational concern: type and predicate
testing~[T], type casting~[C], binding of variables of interest~[B]
and of temporary variables~[b], projection to subelements~[P], success
management~[S], and finally actual reaction~[R].  The design of
Paisley is founded on an object-oriented model and API that separates
these concerns, and reifies their instances as first-class citizens,
as orthogonally as possible.

\begin{figure}
  \begin{lstlisting}[frame=tblr]
public abstract class Pattern<A> {
  public boolean match(A target);
  public boolean matchAgain();
  public Pattern<A> and(Pattern<A> p);
  public Pattern<A> or(Pattern<A> p);
}
  \end{lstlisting}
  \begin{lstlisting}[frame=tblr]
public class Variable<A> extends Pattern<A> {
  private A value;
  public boolean match(A target) {
    value = target;
    return true;
  }
  public boolean matchAgain() {
    return false;
  }
  public A getValue() {
    return value;
  }
}
  \end{lstlisting}
  \begin{lstlisting}[frame=tblr]
public class Motif<A, B> {
  public Pattern<B> apply(Pattern<A> p);
  public <C> Motif<C, B> then(Motif<C, A> m);
  public List<A> eagerBindings(B target);
  public Iterable<A> lazyBindings(B target);
}
  \end{lstlisting}
  \begin{lstlisting}[frame=tblr]
  // miscellaneous
  public static <A> Pattern<A> test(Predicate<A> p);
  public static Pattern<Object> isInstanceOf(Class<?>... c);
  public static Pattern<Object> eq(Object o);
 
  public static <A, B> Motif<A, B> transform(Function<B, A> f);
  public static <A> Motif<A, Object> forInstancesOf(Class<A> c);

  public static <A> Motif<A, A> star(Motif<A, A> m);
  public static <A> Motif<A, A> plus(Motif<A, A> m);
  \end{lstlisting}
  \caption{Paisley core API}
  \label{fig:core}
\end{figure}

The following exposition is a summary of the core layer of Paisley, as
far as required for appreciation of the novel contributions discussed
in the remainder of this paper.  They are presented in synopsis in
Figure~\ref{fig:core}.  The core-level class and method names are
printed in slanted font, to distinguish them from host-level and
user-level items.  Note that the use of generic types has been
simplified for the sake of easy reading.  For more details see
\cite{atps2012}.

Patterns that potentially match target data of type \lstinline|A|
are instances of class \lstinline|Pattern<A>|.  At the
core of the pattern API is its method %
\lstinline|boolean match(A target)|%
.  The type parameter ensures static type safety, to the usual degree
of Java generics.  The Boolean return value indicates success.  Pure
tests~[T] implement \lstinline|match| without side effects, either
manually or by wrapping a predicate object with the static factory
method \lstinline|test|.  There are generic factory methods for type
and equality test, \lstinline|isInstanceOf|, and \lstinline|eq|,
respectively.

Local success management~[S] is implemented by the logical binary
operators \lstinline|and|/\lstinline|or|.  Nondeterminism, such as
introduced by \lstinline|or|, is implemented by explicit backtracking;
after a successful initial match, one may call %
\lstinline|boolean matchAgain()| %
to obtain further solutions, and iterate as long as successful.  Note
that the \lstinline|and| combinator is different from trivial
sequential matching; it is fully distributive over backtracking,
\NEW{analogous to the Prolog \emph{comma} operator.  Operationally, in
  \lstinline|p.and(q)|, \lstinline|q| is (re)started after each
  successful match for \lstinline|p|.  Denotationally, the dependent
  sum of solutions is formed, which degenerates to the Cartesian
  product if \lstinline|q| is independent of the state of
  \lstinline|p|.}

Data flow is by side effects only: Its simplest form is binding [B],
implemented by the subclass \lstinline|Variable<A>| that succeeds
deterministically and binds the matched target data, such that it can
be retrieved if a successful match of the containing pattern has been
performed; otherwise the bound value is undefined.  Temporary
variables~[b] can often be elided by means of point-free pattern
combination.

Transformative data processing, such as type casting~[C] and
projection~[P], is implemented by contravariant lifting; a data
transformation of type \lstinline|B $\to$ A| is lifted to a pattern
transformation of the opposite type %
\lstinline|Pattern<A> $\to$ Pattern<B>|. %
For instance, consider a getter method \lstinline|A getFoo()| of class
\lstinline|B|.  The corresponding lifting is a pattern factory of
roughly the following functionality:
\begin{lstlisting}[basicstyle=\small\sffamily]
  Pattern<B> foo(Pattern<A> p) {
    return new Pattern<B>() {
      public boolean match(B target) {
        return p.match(target.getFoo());
      }
    };
  }
\end{lstlisting}

Pattern factories are again reified in Paisley as class
\lstinline|Motif<A,B>| with methods \lstinline|apply| and
\lstinline|then| for application and composition, respectively, thus
enabling function-level pattern metaprogramming.  The contravariant
lifting operator itself is available as a metafactory method
\lstinline|transform|.  The function to be lifted is typed with a
\emph{functional interface}.  This novel feature of Java~8 emulates
the structural type operator~$(\to)$ in Java's nominal type system,
with automatic coercion to compatible functional interfaces.  Thus
the preceding example can be shortened to the explicit lifting of a
function object,
\begin{lstlisting}
  Motif<A, B> foo = transform(f);
\end{lstlisting}
\NEW{where \lstinline|f| is a method reference, \lstinline|B::getFoo|, or the equivalent lambda expression,
\lstinline|(B b) -> b.getFoo()|,}
which are both significant improvements \NEW{in expressivity} over the pre-Java~8 anonymous
class notation:
\begin{lstlisting}
  Motif<A, B> foo = transform(new Function<B, A>() {
    public A apply(B b) {
      return b.getFoo();
    }
  });
\end{lstlisting}

Corresponding to the type test operator \lstinline|isInstanceOf|, there
is an analogous pattern transformation for casts,
\lstinline|forInstancesOf|.  The pattern transformations of type
\lstinline|Motif<A, A>| form a Kleene algebra with operators
\lstinline|star| and \lstinline|plus|, which we have put to good use
for relational programming \cite{wflp2014}.

In order to apply the Paisley framework to the LISP list data model
from Figure~\ref{fig:pair}, pattern operators for the concrete
API of class \lstinline|Pair| need to be defined.  This model layer of
Paisley is extensible and typically user-defined; the libary provides
only bindings for elementary data, such as classes from the package
\lstinline|java.lang|.
Thus users are free to define their own pattern access style, and to
use the functionality and idioms of the Paisley framework
pragmatically, in whatever way appears most natural and convenient.
Furthermore, since patterns merely bind to the public API of a data
model, no privileged access to source code or runtime objects is
required.  The development of pattern bindings for a data model is
modularly independent from the development of internals of the data
model itself.

Figure~\ref{fig:pair} (right) shows a canonical implementation of
patterns for class \lstinline|Pair|; individual reifications of the
type test and cast operation and getter methods, plus a complex
pattern constructor as the inverse of the data constructor.  Compare
with Figure~\ref{fig:pair} (left).  Figure~\ref{fig:xyz} (bottom)
shows the use of patterns to express the analog of
Figure~\ref{fig:xyz} (center).  A clear separation of concerns as
three sequential statements has been achieved by the basic idiom of
Paisley pattern usage:
\begin{enumerate}
\item allocate pattern variables;
\item construct a complex pattern term that handles testing, casting,
  projection and temporary variables internally and transparently;
\item match, manage success, observe variable bindings and react.
\end{enumerate}

\begin{figure}
  \begin{lstlisting}[frame=tblr]
  Motif<Object, Object> pairCar = asPair.then(car);
  Motif<Object, Object> pairCdr = asPair.then(cdr);
  ^\vspace{-\medskipamount}^
  Motif<Object, Object> nthcdr = star(pairCdr);            ^\Comment{any pair cell}^
  Motif<Object, Object> nth = nthcdr.then(pairCar);        ^\Comment{any element}^
  \end{lstlisting}
  \caption{Enumeration of LISP list elements with Paisley}
  \label{fig:elem}
\end{figure}

In previous applications \cite{wflp2013,wflp2014} we have discussed
how to further encapsulate variable allocation and binding and success
management for encapsulated search over nondetermistic patterns in
logic and functional--logic style, respectively.  For instance,
enumerating the pair cells or elements of a list nondeterministically
is a simple matter of a handful of relational pattern combinators, as
shown in Figure~\ref{fig:elem}.  The class \lstinline|Motif| supports
concise encapsulated search, exposed in collection or iterator style,
via methods \lstinline|eagerBindings| or \lstinline|lazyBindings|,
respectively.  For instance, enumerate all elements of a list as follows:
\begin{lstlisting}
  for (Object e : nth.lazyBindings(list))
    processElement(e);
\end{lstlisting}

In the next section we propose a general, concise, structured
imperative usage style, appropriate for situations where more
fine-grained interleaving of pattern-level and user-level code is
required.

\section{Control for Paisley}

\begin{figure}
  \begin{lstlisting}[frame=tblr]
public class Pattern<A> {
  $\ellipse$
  public Pattern<A> andThen(Runnable r);
  public Pattern<A> orElse(Runnable r);
}
  \end{lstlisting}
  \begin{lstlisting}[frame=tblr]
  public static <A> boolean testThen(A target, Pattern<A> pat, Runnable r) {
    return pat.andThen(r).match(target);
  }
  \end{lstlisting}
  \begin{lstlisting}[frame=tblr]
  public static boolean otherwise(Runnable r) {
    r.run();
    return true;
  }
  public static void ensure(boolean success) {
    if (!success)
      throw new MatchException();
  }
  \end{lstlisting}
  \caption{Continuation extensions in core Paisley}
  \label{fig:runnable}
\end{figure}

The technical contribution of the present paper is an extension of the
Paisley framework, with the purpose of integrating pattern-based
control flow -- case distinctions where clauses are selected not by
plain Boolean conditions but rather by patterns, such that variable
bindings may occur as a side effect and result in data flow to the
selected clause.

Our solution is true to the Paisley principle that object-oriented
programmers should be given declarative expressivity without depriving
them of operational intuition.  Thus, we require a clean separation of
concepts, to the effect that anything that looks like ordinary control
flow of the host language actually behaves as such, and that the
entanglement of control and data flow that is specific to pattern
matching is encapsulated and abstracted appropriately.  We find the
ideal tool for the job in a recent major addition to the Java
language, namely \emph{lambda expressions}.

Consider a more complex variation of the data query example task from
the preceding section, namely to succeed for any list, and distinguish
the four cases of zero, one, two, and three-or-more elements.  Of
course, pattern constructs in analogy to Figure~\ref{fig:xyz} (bottom)
can be used for each case independently.  But the resulting code would
be statically and dynamically redundant, because the common matching
effort is not shared between cases.  The solution is to interleave
user-level success management code with pattern-level testing and
projection code, in a nested fashion at each level of
\lstinline|pair|.

\NEW{Consider Figure~\ref{fig:paircont} (left) ahead for our proposed
  solution style.}  The Paisley binding for the data model is extended
by adding a functional interface representing the \emph{continuation
  transform} of the data class \lstinline|Pair|~(top).  Then a complex
clause operator \lstinline|pairThen|, along the lines of
Figure~\ref{fig:xyz}, which handles data flow and match execution
internally, can be defined~(center).  Its function is to call the
continuation in the event of success, with the data subitems obtained
via pattern variable bindings.  The success flag is returned for
external cascading management.  This yields an idiom for pattern-based
clauses, with Boolean expressions of the form
\begin{lstlisting}
  pairThen(list, (car, cdr) -> {...})
\end{lstlisting}
where the body of the lambda expression that instantiates the
continuation interface is executed as a side effect, if and only if
the query finds that the inverse of the construction
\begin{lstlisting}
  list = new Pair(car, cdr)
\end{lstlisting}
applies.  Additionally, the Boolean expression can be wrapped in a
pattern by means of the \lstinline|test| factory (bottom), for
immediate use
\begin{lstlisting}
  pair((car, cdr) -> {...}).match(list)
\end{lstlisting}
or compositional embedding.  The resulting style is concise and
elegant; data flows directly to well-scoped variables \lstinline|car|
and \lstinline|cdr|, without need to allocate pattern variables and
reason dynamically about their definedness.

Specific continuations, clause operators and pattern wrappers are
user-level constructs specific to a data model.  For their use, only
small extensions to the Paisley core level are required; see
Figure~\ref{fig:runnable}.  We add pattern methods \lstinline|andThen|
and \lstinline|orElse| to affix a parameterless continuation, of
functional interface type \lstinline|Runnable|, in the logically
obvious way.  The generic clause operator \lstinline|testThen|
corresponds to the wrapped form \lstinline|andThen|.  Auxiliary methods
\lstinline|otherwise| and \lstinline|ensure| convert between
\lstinline|boolean| expressions and \lstinline|void| continuations,
respectively, to compensate for the lack of implicit coercions in
Java.

Figure~\ref{fig:xyzcont} explores the space of expressivity of the
contination-based notation.  There are two independent degrees of
freedom, namely the Java-level choice of conditional statements versus
conditional operators, and the Paisley-level choice of Boolean clause
expressions versus wrapped pattern applications.  Thus we obtained a
synopsis of four differently styled but structurally equivalent
solutions to the problem posed above.  We feel that neither of these
styles takes natural precedence, and leave the choice to the taste and
convenience of the user.

If only the success of a pattern matters, but no data flow of
extracted subitems is required, then a simplified query implementation
with parameterless continuations suffices.  The corresponding
operators are depicted in Figure~\ref{fig:paircont} (right).  Note
that the body of \lstinline|pairThen| is defined verbosely only for
the sake of synoptic comparison; it can be constructed more concisely in
a single expression:
\begin{lstlisting}
  testThen(target, isPair, r)
\end{lstlisting}

Omitting unneccessary continuation parameters is beneficial with
respect both to choice of efficient implementation and to temporary
variable hygiene.

\begin{figure}[p]
  \begin{minipage}[t]{0.56\textwidth}
    \begin{lstlisting}[frame=tblr]
interface PairContinuation {
  void cont(Object car, Object cdr);
}
    \end{lstlisting}
  \end{minipage}%
  \setlength\Commentwidth{3cm}%
  \begin{minipage}[t]{0.49\textwidth}
    \begin{lstlisting}[frame=tblr]
interface Runnable { ^\Comment{package java.lang}^
  void run();
}
    \end{lstlisting}
  \end{minipage}\\
  \begin{minipage}[t]{0.56\textwidth}
    \begin{lstlisting}[frame=tblr]
 static boolean pairThen(Object target,
                           PairContinuation pc) {    
   Variable<Object>  car = new Variable<>(),
                     cdr = new Variable<>();
   if (pair(car, cdr).match(target)) {
     pc.cont(car.getValue(), cdr.getValue());
     return true;
   }
   else
     return false;
 }
    \end{lstlisting}
  \end{minipage}%
  \begin{minipage}[t]{0.49\textwidth}
    \begin{lstlisting}[frame=tblr]
 static boolean pairThen(Object target,
                           Runnable r) {  

  
   if (isPair.match(target)) {
     r.run();
     return true;
   }
   else
     return false;
 }
    \end{lstlisting}
  \end{minipage}
  \\
  \begin{minipage}[t]{0.56\textwidth}
    \begin{lstlisting}[frame=tblr]
 static Pattern<Object> pair(PairContinuation pc) {
   return test(x -> pairThen(x, pc));
 }      
    \end{lstlisting}
  \end{minipage}%
  \begin{minipage}[t]{0.49\textwidth}
    \begin{lstlisting}[frame=tblr]
 static Pattern<Object> pair(Runnable r) {
   return test(x -> pairThen(x, r));
 }      
    \end{lstlisting}
  \end{minipage}
  \caption{LISP list continuations (top), clause operators (center) and pattern wrappers (bottom); with data flow (left) and without (right)}
  \label{fig:paircont}
\end{figure}

\begin{figure}[p]
  \noindent
  \begin{minipage}[t]{0.525\textwidth}
    \begin{lstlisting}[frame=tblr]
  if (pairThen(list1, (x, list2) -> {
    if (pairThen(list2, (y, list3) -> {
      if (pairThen(list3, (z, list4) -> {
	case3orMore(x, y, z);
      }));
      else case2(x, y);
    }));
    else case1(x);
  }));
  else case0();
    \end{lstlisting}
  \end{minipage}%
  \begin{minipage}[t]{0.525\textwidth}
    \begin{lstlisting}[frame=tblr]
  if (pair((x, list2) -> {
    if (pair((y, list3) -> {
      if (pair((z, list4) -> {
        case3orMore(x, y, z);
      }).match(list3));
      else case2(x, y);
    }).match(list2));
    else case1(x);
  }).match(list1));
  else case0();
    \end{lstlisting}  
  \end{minipage}
  \\
  \begin{minipage}[t]{0.525\textwidth}
    \begin{lstlisting}[frame=tblr]
  pairThen(list1, (x, list2) -> ensure(
    pairThen(list2, (y, list3) -> ensure(
      pairThen(list3, (z, list4) ->
        case3orMore(x, y, z))
      || otherwise(() -> case2(x, y))))
    || otherwise(() -> case1(x))))
  || otherwise(() -> case0())
    \end{lstlisting}
  \end{minipage}%
  \begin{minipage}[t]{0.525\textwidth}
    \begin{lstlisting}[frame=tblr]
  pair((x, list2) ->
    pair((y, list3) ->
      pair((z, list4) ->
        case3orMore(x, y, z)
      ).orElse(() -> case2(x, y)).match(list3)
    ).orElse(() -> case1(x)).match(list2)
  ).orElse(() -> case0()).match(list1)
    \end{lstlisting}
  \end{minipage}
  \caption{Styles of complex list deconstruction with continuations
    and conditional statements -- with clause operators (left) and
    pattern wrappers (right); as conditional statements (top) and
    expressions (bottom)}
  \label{fig:xyzcont}
\end{figure}

\section{Case Study}

We have applied the Paisley style in general, and the control flow
notation presented in the previous section in particular, to a
real-world, pre-existing, open-source Java project.  We have chosen
the Kawa~2.1\,\cite{kawa} implementation of the Scheme language, a GNU
software package, for this purpose.  The rationale is that complex
queries of a simple data model, namely the LISP lists introduced as
the running example above, is a pervasive topic in Scheme
implementations.

Since we are currently interested in qualitative evaluation, we have
not attempted a full reengineering of the Kawa code.  Instead, we have
exploited the properties of Paisley as a lightweight embedded
domain-specific language, that can be used locally without impact on
the global structure of a program, and experimented with selected
fragments that exhibit typical programming style illustratively.

Figures \ref{fig:example1}, \ref{fig:example2} and \ref{fig:example3}
(left) each show a code fragment from a source file in
\texttt{kawa/standard/}.  Reaction code has been omitted, indicated by
$\ellipse$ markers.  The examples corroborate that our depiction of
the naïve, genuinely object-oriented query style in
Figure~\ref{fig:xyz} is not a parody but standard practice.  The
various concerns are entangled in a way that is highly
non-compositional and obscures the programmer's intentions, hence the
need for a comment.  A notable and typical idiom in this style of
imperative programming is the reuse of temporary reference variables
in queries, and their assignment as a side effect in the midst of
expressions.  We regard the idiom as intensely obfuscated and
error-prone.

The right side of each Figure shows the respective equivalent Paisley
code.  We have aligned all code vertically to highlight the stuctural
correspondence as far as possible, rather than follow the natural
structure of complex Paisley expressions by themselves.  Because
matching for one-element lists is a recurring theme in Figures
\ref{fig:example2} and \ref{fig:example3}, we add a corresponding
clause operator and pattern wrapper to the model-specific library; see
Figure~\ref{fig:singleton}.  This simple expedient results in
considerable code reuse and simplification.

We trust that the direct comparison speaks for itself regarding
simplicity, compositionality and clarity, but name a few points of
particular interest:
\begin{itemize}
\item Recurring patterns can be named mnemonically, simultaneously
  increasing the reuse and the intentional documentation value of
  code, and decreasing the room for errors.
\item Levels of Paisley abstraction can be mixed; for instance, the
  lower third of Figure~\ref{fig:example2} shows a weird query with
  doubly negated continuation, nested within a perfectly regular
  encapsulated search.
\item Figure~\ref{fig:example3} demonstrates that (re)assignment of
  temporary variables can be abstracted away, thus achieving full
  referential transparency, even for complex code.
\end{itemize}

\begin{figure}
  \begin{lstlisting}[frame=tblr]
  static Pattern<Object> triple(Pattern<Object> x, Pattern<Object> y, Pattern<Object> z) {
    return pair(x, pair(y, pair(z, isEmpty)));                                   ^\Comment{cf.\ Figure \ref{fig:xyz}}^
  }
  static Pattern<Object> singleton(Pattern<Object> x) {
    return pair(x, isEmpty);
  }
  static boolean singletonThen(Object target, Runnable r) {
    return testThen(target, pair(any, isEmpty), r);
  }
  \end{lstlisting}
  \caption{Advanced list clause vocabulary}
  \label{fig:singleton}
\end{figure}

Apart from the qualitative and stylistic comparison, we have
investigated simple quantitative criteria; see Table~\ref{tab:quant}.
For each example, we compare the original and the paisleyfied version
with respect to three code metrics:
\begin{enumerate}
\item \emph{The number of lines of code pertaining to matching tasks.}
  Reaction code and context are ignored for this purpose.  Note that
  we have marked some lines of the Paisley version with
  \lstinline|//$\;$*|, to indicate that the line break is solely due
  to alignment with the more verbose original version, and hence
  discounted.
\item \emph{The contribution of matching code to the cyclomatic complexity
  of the method.}  Reaction code and the baseline of one are ignored
  for this purpose.  Note that control-flow branches are moved into
  pattern structure, most notably the logical connectives
  \lstinline|and| and \lstinline|or|.
\item \emph{The number of assignments to temporary variables during
    matching.}  We regard a variable as temporary only if it is not
  observed directly by reaction code, and count assignments rather
  than declarations to account for the variable-reusing style of the
  original code.  Parameters of Paisley continuations count as implied
  assignments.
\end{enumerate}
We find significant improvements for all three metrics, even if the
meaningfulness of lines of code and cyclomatic complexity, both
statement-centric metrics, are somewhat diminished in the more
expression-centric Paisley style.

\begin{table}
  \caption{Quantitative assessment of case study examples}
  \label{tab:quant}
  \begin{center}
    \small
  \begin{tabular}[tabular]{lrrrrrrrrr}
    \toprule
    \multicolumn{1}{c}{\bfseries Example} & \multicolumn{3}{c}{\bfseries Lines of Code} & \multicolumn{3}{c}{\bfseries Cyclomatic Complexity} & \multicolumn{3}{c}{\bfseries Temporary Assignments}
    \\
    & original & Paisley & \em saving 
    & original & Paisley & \em saving
    & original & Paisley & \em saving
    \\ \midrule
    \ttfamily export & 18 & 10 & \em 44\% & 7 & 1 & \em 86\% & 6 & 3 & \em 50\%
    \\
    \ttfamily module\_static & 26 & 17 & \em 35\% & 14 & 3 & \em 79\% & 7 & 4 & \em 43\%
    \\
    \ttfamily IfFeature & 28 & 16 & \em 43\% & 12 & 2 & \em 83\% & 10 & 3 & \em 70\%
    \\ \bottomrule
  \end{tabular}
\end{center}
\end{table}

\section{Conclusion}

We have demonstrated how to implement fine-grained integration of
Paisley pattern-matching logic and data flow with Java control flow,
using the novel lambda expressions to encode the right-hand sides of
pattern clauses as continuations.  This achievement greatly widens the
scope of applicability of Paisley to more complex situations.

With the Kawa case study, we have demonstrated that the Paisley
approach solves a practically relevant problem, and can be used
effectively to reengineer legacy code in a local and incremental
fashion.  This is possible because Paisley is lightweight by design,
and abstracts only from operational low-level burdens of matching
\emph{procedures}, without imposing alternative, declarative semantics
on data and pattern \emph{objects}.

\subsection{Related Work}

The most comprehensive attempt to combine the Java language with
nondeterministic pattern matching is JMatch\,\cite{jmatch}.  Their
approach is both semantically more ambitious, adding declarative
features to the Java type system in the form of so-called \emph{modal
  interfaces}, and technologically more heavyweight, requiring
implementation mechanisms such as coroutines and continuation-passing
style transforms; thus it is implemented as a proprietary language
extension.

In \cite{jmatch-patterns} they have reported reductions of program
complexity similar to our present findings.  Because of their
intrusion into the type system, it is possible to reason statically
about the totality of JMatch pattern clauses, whereas our approach
shares the fate of most imperative object-oriented code, and can only
be feasibly verified informally by inspection and testing.  On the
other hand, because of the heavy-handed implicit transformation of
JMatch programs, the possible interferences with other language
features such as non-local control flow, concurrency, instrumentation
and debugging are unclear.  Furthermore, despite recent award-winning
theoretical publications \cite{jmatch-patterns}, their implementation
is stuck with an obsolete snapshot of the host language Java\,1.4
(sic!), and hence of no practical relevance for contemporary software
production.

\subsection{Outlook}

Our treatment of pattern matching in conditional control flow enforces
deterministic use: the backtracking method \lstinline|matchAgain| is
not invoked explicitly during clause selection. Only the first
solution of a pattern is used, and the continuation is accordingly
called at most once.  However, this does not preclude \emph{internal}
nondeterminism; subpatterns may have to backtrack in order to find
that solution.  As a future generalization, we consider both loop-like
(eager) and iterator-like (lazy) control-flow notations for
\emph{reentrant} continuations selected by nondeterministic patterns.

The refactoring changes to the Kawa code base do not produce
sufficient regular dynamic coverage in order to evaluate the
performance of Paisley pattern matching in relation to naïve
implementations.  We are looking forward to other case studies where
data querying is more predictibly dominant in program running times.
In particular, we are curious about the capability of just-in-time
compilers to optimize `hot' Paisley code in realistic scenarios.

As a final optimistic observation, Paisley has proven resilient to
evolution of the host language, in marked contrast to JMatch, because
it is lightweight and thus syntactically and semantically reducible.
We have anticipated the use of functional interfaces, based on
common-sense extrapolation from Pizza and Scala, and are now
immediately ready to reap the benefits of lambda expressions.
Whatever the new features of Java 9 and 10 will be, the odds are that
Paisley can profit.

\setlength\Commentwidth{1.5cm}

\begin{sidewaysfigure}
  \begin{minipage}[t]{0.51\textwidth}
    \begin{lstlisting}[frame=tblr,basicstyle=\small\sffamily]
Object symbol; Translator tr;                                ^\Comment{context}^





 if (symbol instanceof Pair) {
   // Match (rename name1 name2)
   Pair psym = (Pair) symbol;
   Object symcdr, symcddr; Pair psymcdr;
   if (tr.matches(psym.getCar(), "rename")
       && (symcdr = psym.getCdr()) instanceof Pair
       && ((symcddr = (psymcdr = (Pair) symcdr)
            .getCdr()) instanceof Pair)) {
       Pair psymcddr = (Pair) symcddr;
       Object symcdddr = psymcddr.getCdr();
       Object name1 = tr.namespaceResolve(psymcdr.getCar());
       Object name2 = tr.namespaceResolve(psymcddr.getCar());
       if (symcdddr == LList.Empty
           && name1 instanceof Symbol
           && name2 instanceof Symbol) {
         $\ellipse$
       } 
    }
 }    
    \end{lstlisting}
  \end{minipage}%
  \begin{minipage}[t]{0.49\textwidth}
    \begin{lstlisting}[frame=tblr,basicstyle=\small\sffamily]
Object symbol; Translator tr;                             ^\Comment{context}^

 Variable<Symbol>  vname1 = new Variable<>(), 
                   vname2 = new Variable<>();
 Motif<Symbol, Object> nr = transform(x -> tr.namespaceResolve(x))
                            .then(forInstancesOf(Symbol.class));
 if (triple(                                              ^\Comment*^



            test(x -> tr.matches(x, "rename")),
            nr.apply(vname1),
            nr.apply(vname2)).match(symbol)) {



   Symbol name1 = vname1.getValue();
   Symbol name2 = vname2.getValue();



     $\ellipse$


 }
    \end{lstlisting}  
  \end{minipage}
  \caption{Example code fragment
    (\texttt{kawa/standard/export.java}, lines 52ff.) in Kawa
    original (left) and Paisley style (right)}
  \label{fig:example1}
\end{sidewaysfigure}

\begin{sidewaysfigure}
  \begin{minipage}[t]{.55\textwidth}
    \begin{lstlisting}[frame=tblr,basicstyle=\footnotesize\sffamily]
Pair st; Translator tr;                                                        ^\Comment{context}^

    Object list = st.getCdr();
    $\ellipse$
    if (list instanceof Pair
         && (st = (Pair) list).getCdr() == LList.Empty
         && st.getCar() instanceof Boolean) {
        if (st.getCar() == Boolean.FALSE)
            $\ellipse$
        else
            $\ellipse$
    } else if (list instanceof Pair
                && (st = (Pair) list).getCdr() == LList.Empty
                && st.getCar() instanceof Pair
                && tr.matches((st = (Pair) st.getCar()).getCar(),
                              Scheme.quote_str)) {
        Object cdr = st.getCdr();
        if (cdr != LList.Empty
             && (st = (Pair) cdr).getCar() instanceof SimpleSymbol
             && st.getCar().toString() == "init$\text{-}$run") {                         ^\Comment{sic!}^
            $\ellipse$
        } else {
            $\ellipse$
        }
    } else {


        while (list != LList.Empty) {
            if (! (list instanceof Pair)
                 || ! ((st = (Pair) list).getCar() instanceof Symbol)) {
                $\ellipse$
            } else {
                $\ellipse$ // uses (st, (Symbol)st.getCar())
            }
            list = st.getCdr();
        }
    }
    \end{lstlisting}
  \end{minipage}%
  \begin{minipage}[t]{0.45\textwidth}
    \begin{lstlisting}[frame=tblr,basicstyle=\footnotesize\sffamily]
Pair st; Translator tr;                                      ^\Comment{context}^

    Object list = st.getCdr();
    $\ellipse$
    if (singleton(                                           ^\Comment*^


                  eq(Boolean.FALSE))
        .andThen(() -> $\ellipse$)
        .or(singleton(eq(Boolean.TRUE))
            .andThen(() -> $\ellipse$))
        .or(singleton(                                       ^\Comment*^

                pair(
                    test(x -> tr.matches(x,                  ^\Comment*^
                                         Scheme.quote_str)),

                    pair(
                        test(x -> x instanceof SimpleSymbol
                                  && x.toString() == "init$\text{-}$run"),
                        any).andThen(() -> $\ellipse$)
                    .orElse(() -> {
                        $\ellipse$
                    })))).match(list));
    else {
        Variable<Symbol> symbol = new Variable<>();
        Variable<Pair> p = new Variable<>();
        for (Object r : nthcdr.lazyBindings(list)) {
            if (!asPair.apply(car.apply(asSymbol(symbol))
                              .and(p)).match(r)) {
                    $\ellipse$
            } else {
                $\ellipse$ // uses (p.getValue(), symbol.getValue())
            }

        }
    }
    \end{lstlisting}
  \end{minipage}
  \caption{Example code fragment
    (\texttt{kawa/standard/module\_static.java}, lines 23ff.) in Kawa
    original (left) and Paisley style (right)}
  \label{fig:example2}
\end{sidewaysfigure}

\begin{sidewaysfigure}
  \begin{minipage}[t]{0.5\textwidth}
    \begin{lstlisting}[frame=lrbt,basicstyle=\footnotesize\sffamily]
Object form; Translator tr;                                          ^\Comment{context}^

    form = tr.namespaceResolve(Translator.stripSyntax(form));
    if (form instanceof String || form instanceof SimpleSymbol) {
        $\ellipse$
    }
    if (form instanceof Pair) {
        Pair pair = (Pair) form;
        Object keyword = Translator.stripSyntax(pair.getCar());
        if (keyword == orSymbol || keyword == andSymbol) {
            Object rest = pair.getCdr();
            while (rest instanceof Pair) {
                pair = (Pair) rest;
                $\ellipse$
                rest = pair.getCdr();
            }
            $\ellipse$
        }
        if (keyword == notSymbol) {
            Object rest = pair.getCdr();
            if (rest instanceof Pair) {
                Pair pair2 = (Pair) rest;
                if (pair2.getCdr() == LList.Empty)
                    $\ellipse$
            }
            $\ellipse$
        }
        if (keyword == librarySymbol) {
            Object rest = pair.getCdr();
            if (rest instanceof Pair) {
                Pair pair2 = (Pair) rest;
                if (pair2.getCdr() == LList.Empty)
                    $\ellipse$
            }
            $\ellipse$
        }
    }
    \end{lstlisting}
  \end{minipage}%
  \begin{minipage}[t]{0.5\textwidth}
    \begin{lstlisting}[frame=lrbt,basicstyle=\footnotesize\sffamily]
Object form; Translator tr;                                          ^\Comment{context}^


    isInstanceOf(String.class, SimpleSymbol.class).andThen(() ->  {
        $\ellipse$
    })
    .or(pair((keyword, rest) -> {


        eq(orSymbol).or(eq(andSymbol)).andThen(() ->  {

            for (Pair pair : nthcdr.then(asPair).lazyBindings(rest)) {

                $\ellipse$

            }
            $\ellipse$
        })
        .or(eq(notSymbol).andThen(() -> {

            singletonThen(rest, () -> {


                    $\ellipse$
            });
            $\ellipse$
        }))
        .or(eq(librarySymbol).andThen(() -> {

            singletonThen(rest, () -> {


                    $\ellipse$
            });
            $\ellipse$
        })).match(Translator.stripSyntax(keyword));
    })).match(tr.namespaceResolve(Translator.stripSyntax(form)));
    \end{lstlisting}
  \end{minipage}
  \caption{Example code fragment
    (\texttt{kawa/standard/IfFeature.java}, lines 42ff.) in Kawa
    original (left) and Paisley style (right)}
  \label{fig:example3}
\end{sidewaysfigure}

\renewcommand\markboth[2]{\relax}
\renewcommand\bibfont{\small}
\printbibliography

\end{document}